\documentclass[aps,nofootinbib,prd,superscriptaddress,reprint]{revtex4-1}
\usepackage{amssymb}
\usepackage{amsmath}
\usepackage{amsfonts}
\usepackage{graphicx,color}
\usepackage{mathrsfs}
\usepackage{hyperref}

\newcommand{\kets}[1]{\left\vert #1 \right\rangle}
\newcommand{\bras}[1]{\left\langle #1 \right\vert}
\newcommand{\beq}{\begin{equation}}
\newcommand{\eeq}{\end{equation}}
\newcommand{\bqr}{\begin{eqnarray}}
\newcommand{\eqr}{\end{eqnarray}}

\begin{document}

\author{Sujoy K. Modak}
\email{sujoy.kumar@correo.nucleares.unam.mx}
\author{Leonardo Ort\'{i}z}
\email{leonardo.ortiz@correo.nucleares.unam.mx}
\affiliation{Instituto de Ciencias Nucleares, Universidad Nacional Aut\'onoma de M\'exico, M\'exico D.F. 04510, M\'exico}
\author{Igor Pe\~na}
\email{igor.pena@uacm.edu.mx}
\affiliation{Plantel Casa Libertad, Universidad Aut\'{o}noma de la Ciudad de M\'{e}xico, Calzada Ermita Iztapalapa 4163, Distrito Federal, 09620, M\'{e}xico}
\author{Daniel Sudarsky}
\email{sudarsky@nucleares.unam.mx}
\affiliation{Instituto de Ciencias Nucleares, Universidad Nacional Aut\'onoma de M\'exico, M\'exico D.F. 04510, M\'exico}

\title{  Non-Paradoxical Loss  of  Information in Black Hole  Evaporation  in a Quantum Collapse Model}

\begin{abstract}
We consider  a novel  approach  to address the black hole information paradox (BHIP). The idea is based  on adapting, to the situation  at  hand,   the modified  versions of quantum theory  involving spontaneous stochastic    dynamical collapse  of quantum states, which have  been considered  in attempts to  deal with    shortcomings  of  the  standard Copenhagen interpretation of quantum mechanics,  in particular, the issue  known as ``the measurement problem". The    new  basic  hypothesis  is that  the modified  quantum behavior  is  enhanced  in the region of high curvature  so that  the information encoded  in the  initial quantum  state  of  the matter fields  is rapidly   erased as  the black hole singularity  is  approached. We show that in this manner the complete   evaporation of   the black hole   via  Hawking radiation  can  be  understood  as involving no paradox.  Calculations are performed  using a  modified  version of quantum theory known as ``Continuous  Spontaneous Localization"   (CSL), which was  originally  developed  in the context of  many particle non-relativistic  quantum mechanics. We  use  a version of CSL tailored to  quantum field theory  and   applied in the context of the two  dimensional Callan-Giddings-Harvey-Strominger (CGHS) model. Although the  role  of quantum gravity  in this picture  is  restricted  to  the resolution of the singularity,  related  studies  suggest  that  there    might be    further  connections.
\end{abstract}

\maketitle

\section{Introduction}

The essence of  the  information loss paradox in black hole (BH) evaporation is that, starting  with a system in a pure initial quantum state  of matter that  forms   a  BH,  its     subsequent evolution   leads   to something that, at the quantum level,  can only be characterized  as a  highly  mixed  quantum state,  while,  the standard  quantum  mechanical considerations   lead one  to expect a fully unitary evolution \cite{hawk2}. There is  even a debate   as to whether or not  this  should be considered  as a paradox \cite{Wald-book}. In \cite{Okon-New} it  is argued  that the  debate    arises  due  to  basic  differences of outlook  regarding  the fate  of the  BH singularity  in the context of  a quantum  theory.

 In recent times,  many  researchers  in the  quantum gravity (QG)   field have  been  trying  to address  the   black hole information conundrum, within the context  of   their preferred  approach. After all,  the ``paradox"  truly  emerges only  if one   assumes that QG will  remove the  singularities  that   appear in association with black holes  in   general  relativity (GR).  Otherwise,   the singularity   could be  viewed as  representing  an additional  boundary of space-time,  where the missing  information could be registered. The proposals to address the issue  involve  various schemes  whereby the  complete evolution  respects   quantum mechanical  unitarity, and  thus the information   is    strictly conserved \cite{maldacena}-\cite{Mathur2}. However, whereas there are questions on the validity of some of these proposals \cite{bojowald}, other studies along those lines  \cite{FireWalls} connect the resolution of the issue  with  the emergence  of  ``firewalls'',  creating  a serious tension between the equivalence principle of GR and the unitarity of quantum mechanics.
The  considerations  involving ``firewalls"  have   induced  an  intense controversy as to which of the two  basic tenets,  the unitarity  of  the  evolution  or  the equivalence principle,  should be sacrificed in  resolving the issue.
It  is  clear  that none  will be without   repercussions  either in  quantum mechanics  (QM), or  in the  theory of   general relativity.
Presently, there  are  various dramatic  ideas  under  consideration.  For instance,  that the event horizon never forms \cite{no-hor},  that  even if there  is  an event horizon, particles inside and outside  the horizon are entangled via wormholes (ER=EPR hypothesis) \cite{er=epr}, etc.  We do not find   these  proposals   very   attractive,  among other  reasons,   because  they  have   been designed,  exclusively,   to  deal   with   the   problem at   hand and  seem to   lack a   ``broader" motivation.
 In contrast,  the  proposal   we   will explore in  this  work,   connects the   issue  at hand to  what is often taken  as the most serious  foundational problem in quantum theory,  namely ``the measurement problem".
 We have  been motivated  by Penrose's  long standing arguments \cite{Penrose}  suggesting that  the   reconciliation of quantum theory and   general relativity,  something that is often  considered  as  limited to  finding  an appropriate theory of QG ,  will  in fact require much more, i.e.  the modification of both  QM and  GR, and by his observation that a statistical
 picture of thermal  equilibrium  of  a system,  which includes  black holes,  seems  to  require some departure from   unitarity in the evolution of  ordinary quantum systems \cite{Penrose-BH-Collpase}.

 The overall scheme  we  will consider is  based on  the   dynamical  reduction
 theories   introduced to deal  with the  ``measurement problem" \cite{measurement}-\cite{Mau:95}\footnote{ We recall that the issue here is  how to   make  compatible the   reduction postulate  associated  with measurements  with  the    dynamical evolution   law provided  by Schr\"odinger's  equation if one wants to describe   the apparatus and observer  and  not just the system of interest,  in terms of the quantum theory.}.
These  ideas date back to  \cite {Bhom-Bub}, with  the first specific toy model  proposals in \cite {Pearle:7679}, and the first viable proposals  in \cite{GRW:8586}  with the theory of Spontaneous Localization, and  later  the proposal   known as  Continuous  Spontaneous  Localization or CSL \cite{Pearle:89, GPR:90}. Not long after that,  Diosi  \cite{diosi} and Penrose \cite{Penrose-BH-Collpase, Penrose}  proposed the  connection   of these ideas  with   quantum  aspects of gravity.   Recently, these  ideas    have been   considered by Weinberg \cite{Weinberg:2011jg},  who emphasizes  that,  despite all  efforts,  we still  lack  a  reasonable  interpretational  solution to  the problems of quantum theory.
 We will   consider  the issue at hand  within  the context of   such theories, and   ask  if   they could account,  at the  quantitative level, for  the loss of information in black hole evaporation,  and   about the kind  of adaptations that  would be involved  in doing so.

   These  theories   include  modifications of quantum dynamics which  can be   characterized  by    additional   terms in    the Schr\"odinger  equation.  The   impact  of such terms  is controlled  by a new  fundamental  constant called the collapse parameter, 
   often denoted  by $ \lambda$, which is  taken  to  be  small enough  to ensure that  
   ordinary quantum mechanics  holds  as  a very  good  approximation in regimes  not 
   involving  too  many  particles (where it has been tested  with enormous precision), and  
   yet large  enough to   explain the  absence of macroscopic  Schr\"odinger cat  states.
 We make the novel  hypothesis that  $ \lambda$ {\it  depends   strongly  
  on  the space- time curvature}.
 That is, in the regions of  smaller or   vanishing  curvature,  this ``stochastic collapse" 
 will be negligible, and effective quantum evolution will  be given essentially  
  by Schr\"odinger's evolution (except  when large  apparatuses consisting of  too many particles are involved, as in the cases considered  by the original collapse theories). But, in regions of higher curvature,
 the evolution will be dominated by this  new term,  making the effective evolution {\it  highly non-unitary}.
 We  should  note here the   connections   between the   present proposal and the  arguments  regarding  the essential  viability of models  involving   loss of unitarity in the effective  description of a black  hole  evaporation, which  were   carried  out in \cite{Unruh-Wald}.

It should  be stressed that resolving the BH information paradox  within  this  kind of  approach would require explaining how a pure state  becomes  a (quantum) thermal state corresponding to a \emph{proper} mixture, rather than an \emph{improper} one\footnote{A proper  mixture represents an actual ensemble of  systems,  each of   which has been prepared  to be in different but  definite  states,  with  their   proportion in the  ensemble   determined  by  the corresponding   weights.
An improper mixture represents  the partial  description,   as  provided  by the reduced  density matrix,  of a  subsystem  which is  part  of a larger  system which is,  as a whole,  in a pure state.  This terminology  is borrowed  from \cite{dEspagnat, timpson}.},  as a result  of  the  eventual    disappearance of  the interior  region,   in a  the case  of complete   evaporation of the black hole.
  We will see this  at work  in the analysis  below.

\section{The Basic Set up}

   Given the  complexity of the problem we  will   present our analysis using a simplified   2-dimensional model known as the  Callan-Giddings-Harvey-Strominger (CGHS)  black hole \cite{CGHS92}, and   will   work  with  a  toy- adapted version  of  CSL.

  The   modified  quantum  dynamical   evolution,  as dictated by   CSL, is specified by the choice of a  certain observable $\hat A$, and by two equations:
  i) A  (stochastically) modified Schr\"odinger equation, whose  solution is:
  \begin{equation}\label{CSL1}
 { |\psi,t\rangle_w = \hat {\cal T}e^{-\int_{0}^{t}dt'\big[i\hat H+\frac{1}{4\lambda_{0}}[w(t')-2\lambda_{0}\hat A]^{2}\big]}|\psi,0\rangle,}
\end{equation}
 where $\hat {\cal T}$ is the time-ordering operator,  $w(t)$ is a random, white noise type classical function of time whose probability is given by the second equation,  ii) the Probability Distribution Density $[P(Dw(t))]$ rule:
  \begin{equation}\label{CSL2}
	 { P(Dw(t))\equiv{}_w \langle\psi,t|\psi,t\rangle_w \prod_{t_{i}=0}^{t}\frac{dw(t_{i})}{\sqrt{ 2\pi\lambda_{0}/dt}}}.
\end{equation}
Thus  the  standard   Schr\"odinger    evolution   and the changes in the state corresponding  to a ``measurement" of  the observable {$\hat A$} are  unified. For  non-relativistic   quantum  mechanics   of a single particle,  the proposal assumes that  there is,  in all situations (without invoking any  measurement device  or observer),  a    spontaneous and continuous  reduction  characterized  by   {$\hat A = \hat {\vec X}_\delta $}, where  $\hat{\vec X}_\delta$ is  a suitably  smeared   (with the  smearing characterized by  the scale $\delta$) version of the position operator  $\hat{\vec X}$.   When this  is  generalized to  multi-particle  systems\footnote{This,  in particular,  involves dealing with multiple  operators $\hat{A}$  (one for each particle),  and their corresponding stochastic functions $w$.  The  evolution equation then takes the form  we  will use in  \ref{csl-ev}. } and  everything,  including, the  apparatuses are  treated  quantum mechanically, the theory  seems  to   successfully address the   measurement problem \cite{measurement}-\cite{Mau:95}. For   all this   to work appropriately  at the quantitative level, the  collapse  parameter  {$\lambda_{0}$} must be  small enough not to  conflict with tests of QM  in the domain of   subatomic physics,  and big enough  to result in rapid localization of   ``macroscopic objects".   The GRW  suggested  value is {$\lambda_{0} \sim 10^{-16}  sec ^{-1}$}.  For more discussions  regarding  the  current status of the theory, and in particular, the  empirical  constraints,  we refer the reader to \cite{More-CSL}.


So far; studies related to CSL dynamics have been mostly  limited  to non-relativistic   many particle systems   in flat space-time\footnote{We should also  note the  recent  investigations   involving  application of CSL to the  problem of   generation of the seeds of  cosmic  structure in inflationary cosmology \cite{CSL-cosmology, pedro}.}.  We must, however,  beware  of trying, at this  stage, to compare those studies  with the present one, and  to  require, for instance that the exact versions, of the theory  used in  the   vastly different contexts, coincide. In that sense,  we must view the present stage of investigation of  these  ideas   as  a  search for the  basic  clues that  will  permit  the   delineation of  a  fundamental collapse theory,   which   should   not  only be applicable in  all situations, (i.e., cosmology,  black holes  and laboratory setting),   but  should also be clearly  self consistent and  fully covariant. In the present work,  we must   adapt the approach to  situations  involving  both  quantum fields  and curved space-times.

We note that dynamical reduction in the quantum state  requires the notion of  ``time" (the collapse takes place in time), and given that   canonical QG   is known to  have a problem with time, we   will proceed  with  our  analysis assuming that,  to a large  extent, and  in fact   in regimes   where  the   curvature  is   far from the Planck scale,   one   can rely on  the  semi-classical  framework.  Our point of view  is   that, even if  at the  fundamental  level  gravitation  must be  quantum mechanical  in nature, at the   sub-Planckian  scales,  it  should be  described   in  terms of   semi-classical gravitation,   where the metric  itself corresponds  to  an  emergent phenomena,  and where traces of the  full quantum regime might  survive and   include  effects which, at in  semi-classical level of  description,  take  the form of  an  effective  dynamical  state reduction for matter fields.

To summarize, our analysis is based on following ingredients:
\begin{enumerate}
   \item  The   CGHS black hole,
   \item a toy version of CSL  adapted to a field theory on a curved space-time,
  \item an assumption that the  CSL  collapse parameter is not  fixed  but  depends (increases)  with  the  local curvature \cite{noparadox} and
   \item some   simplifying, but  rather natural,  assumptions  about   what  happens   when  QG ``cures"  a singularity.
  \end{enumerate}

\section{CSL theory in CGHS Model}

\subsection{CGHS model}

\begin{figure}
\includegraphics[scale=0.35]{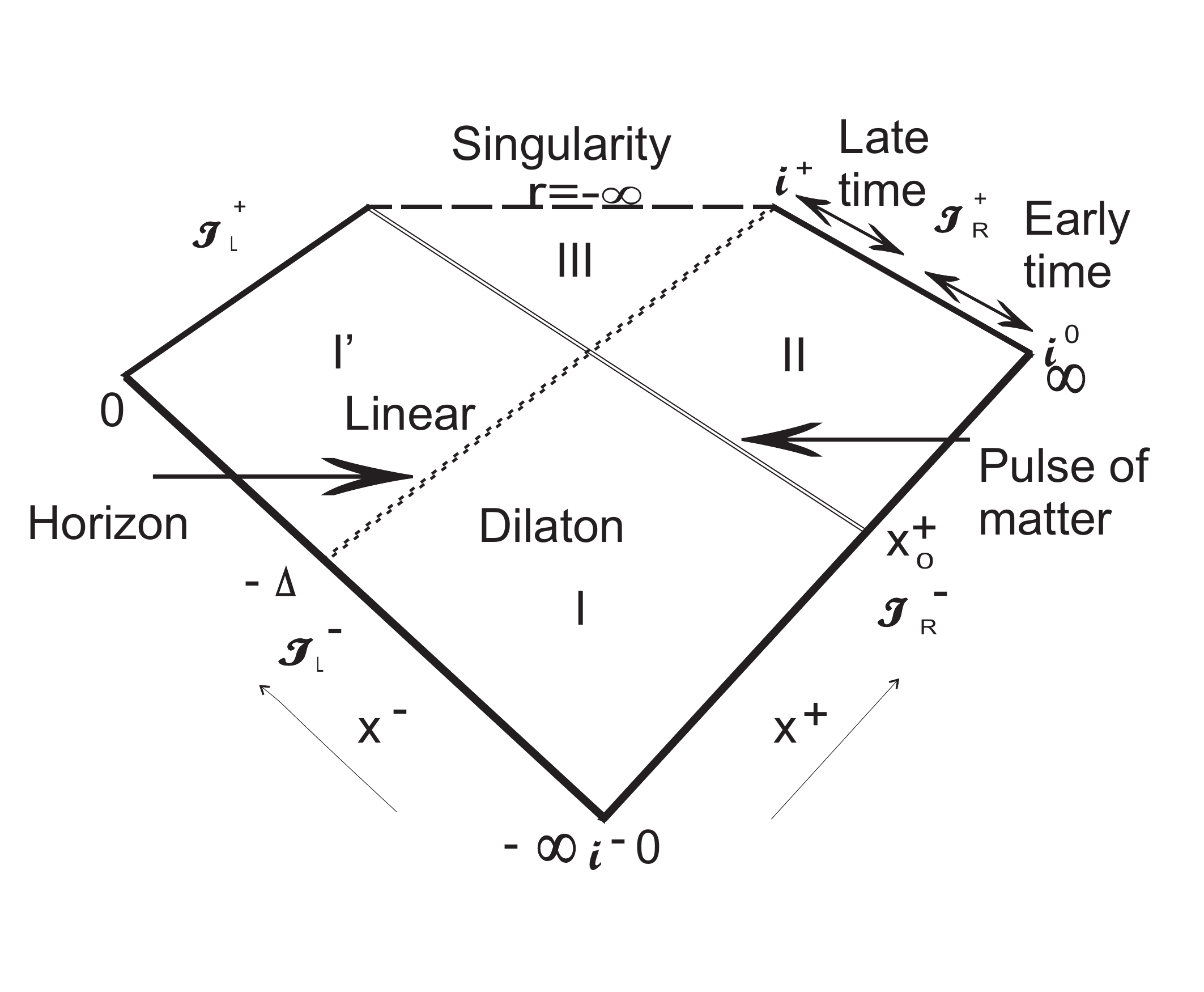}
\caption{Penrose diagram for CGHS spacetime.}
\label{cghs}
\end{figure}

Now, we review   the basic  features  of the  CGHS model, which offers  a two dimensional version  of black hole formation   and evaporation. For more details we refer the reader to \cite{cghs-more}. The CGHS action is given by
\beq
{ S=\frac{1}{2\pi}\int d^2x\sqrt{-g}\left[e^{-2\phi}\left[R+4(\nabla \phi)^2+4\Lambda^2\right]-\frac{1}{2}(\nabla f)^2\right]} \nonumber
\eeq
where { $\phi$} is the dilaton field, { $\Lambda^2$} is the cosmological  constant, and { $f$} is a  scalar  field,  representing  matter. The generic construction of the CGHS model is shown in Fig. \ref{cghs}.
At $x^+ < x^+_{0}$, the metric is Minkowskian, usually known as the dilaton vacuum (region I and I'), given by ${ds^2=-\frac{dx^{+}dx^{-}}{-\Lambda^2 x^{+}x^{-}},}$ whereas, at $x^+ > x^+_0$ it is represented by the black hole metric (region II, III) $ds^2=-\frac{dx^{+}dx^{-}}{\frac{M}{\Lambda}-\Lambda^2 x^{+}(x^{-}+\Delta)}.$ These  (null) Kruskal  coordinates ($x^+, x^-$) are useful in presenting the global structure of the spacetime. On the other hand, for physical studies involving Quantum Field Theory (QFT) in curved space-time, it is often convenient to use  coordinates associated   with physically interesting observers in  various regions. In the dilation vacuum  region, natural Minkowskian coordinates are $y^+ =\frac{1}{\Lambda} \ln(\Lambda x^+) , y^- = \frac{1}{\Lambda} \ln(-{\frac{x^-}{\Delta}})$, and the metric is expressed as $ds^2=-dy^{+}dy^{-}$ with $-\infty < y^{-} <\infty;~ -\infty <y^{+} < \frac{1}{\Lambda}\ln(\Lambda x_0^+)$. On the other hand, on the BH  exterior (region II), where  the long lived physical observers might  exist, one has  the coordinates $\sigma^+ = \frac{1}{\Lambda} \ln(\Lambda x^+) = y^+, \sigma^- = -\frac{1}{\Lambda} \ln(-\Lambda(x^- + \Delta))$, so that the metric is $ds^2=-\frac{d\sigma^{+}d\sigma^{-}}{1+(M/\Lambda)e^{\Lambda(\sigma^{-}-\sigma^{+)}}}$ with $-\infty < \sigma^{-} < \infty$ and $\sigma^{+} > \sigma_{0}^{+}= \frac{1}{\Lambda} \ln(\Lambda x_0^+).$ We can see  the asymptotic flatness by expressing this  metric in Schwarzschild like coordinates {($t,r$)} defined by {$\sigma^{\pm} =t \pm \frac{1}{2\Lambda} \ln(e^{2\Lambda r} - M/\Lambda)$} so that, we have ${ ds^2= -(1-\frac{M}{\Lambda}e^{-2\lambda r}) dt^2 + \frac{dr^2}{(1-\frac{M}{\Lambda}e^{-2\lambda r})}}$. The Kruskal coordinates $2T =  x^+ + x^- + \Delta,~2X =x^+ - x^- -\Delta$ can be related with Schwarzschild like time $t$ and space $r$ coordinates using $\tanh(\Lambda t) = T/X$ and $-\frac{1}{\Lambda^2}(e^{2\Lambda r} - M/\Lambda) = T^2-X^2$.

  Next,  we consider  the  quantum description of the field  $f$  for which one   uses  
   ${\mathscr{I}}^{-}_{L} \cup {\mathscr{I}}^{-}_{R}$ as the asymptotic  past  ({\it in}) region,  and the black hole (exterior and interior) region as the asymptotic {\it out} region. The {\it in} description of  the quantum  field operator can be expanded as $\hat{f}(x) = \sum(\hat{f}_{\omega}^{R}(x) + \hat{f}_{\omega}^{L}(x)),$ where $f_{\omega}^{R/L} = \hat{a}_{\omega}^{R/L} u_{\omega}^{R/L} + \hat{a}^{R/L \dag}_{\omega} u_{\omega}^{R*/L*}$.
Here, the basis of functions (modes) are:
$
{u_{\omega}^{R}=\frac{1}{\sqrt{2\omega}}e^{-i\omega y^{-}}}
$
and
$
{u_{\omega}^{L}=\frac{1}{\sqrt{2\omega}}e^{-i\omega y^{+}}},
$
with {$\omega>0$}. The superscripts {$R$} and {$L$} mean right and left moving modes.
These modes  thus specify a right {\it in} vacuum ({$\kets{0_{in}}_{R}$}), and a left {\it in} vacuum ({$\kets{0_{in}}_{L}$}),
whose tensor product ({$\kets{0_{in}}_{R}\otimes \kets{0_{in}}_{L}$}) defines our {\it in} vacuum. One can also expand the field in the {\it out} region in terms of the complete set of modes that (at late times)   have support in the  outside (exterior) and inside (interior) of the event horizon, respectively. Therefore the field operator has the form $\hat{f}(x) = \hat{f}_{out}^{R}(x) + \hat{f}_{out}^{L}(x)$ where,
\bqr \nonumber
\hat{f}_{out}^{R/L}(x) &=& \sum_{\omega}\hat{b}_{\omega}^{R/L} v_{\omega}^{R/L} +\qquad \hat{b}^{R/L \dag}_{\omega} v_{\omega}^{R*/L*} \\
&& + \sum_{\tilde\omega}\hat{b}_{\tilde{\omega}}^{R/L} v_{\tilde{\omega}}^{R/L} + \hat{b}^{R/L \dag}_{\tilde{\omega}} v_{\tilde{\omega}}^{R*/L*},
\eqr
where we use the convention in which modes with and without tildes are associated with    having  support  inside and outside the horizon, respectively. The  corresponding operators are   similarly labeled.  We note   that ,  as  will be further argued in the   last section, the  arbitrariness  in the choice of  modes inside the horizon will   not affect  our physical results. The   mode  functions  in the exterior to the  horizon  that  we  will use are:
$
{v_{\omega}^{R}=\frac{1}{\sqrt{2\omega}}e^{-i\omega\sigma^{-}}\Theta(-(x^{-} + \Delta))}$
and
$
{v_{\omega}^{L}=\frac{1}{\sqrt{2\omega}}e^{-i\omega\sigma^{+}}\Theta(x^{+} - x_0^{+}).}$
Similarly, one can define a set of modes  in the black hole interior  so that the basis of modes in the {\it out} region is complete.  For the left moving modes,  we  maintain the same  choice as before, while  for  the right moving mode, we use
$
{\hat{v}_{\tilde{\omega}}^{R}=\frac{1}{\sqrt{2\tilde{\omega}}}e^{i\tilde{\omega}\sigma_{in}^{-}}
\Theta(x^{-} + \Delta)}
$.
  Following \cite{cghs-more}, we replace the above delocalized plane wave modes by a complete orthonormal set of discrete wave packets modes, $
v_{nj}^{L/R}=\frac{1}{\sqrt{\epsilon}}\int_{j\epsilon}^{(j+1)\epsilon}d\omega e^{2\pi i\omega n/\epsilon}v_{\omega}^{L/R},
$where the integers $j\ge 0$ and  $-\infty <n<\infty$. These wave packets are peaked about $\sigma^{+/-}= 2\pi n/\epsilon$ with width $2\pi/\epsilon$  respectively.

   The non-trivial Bogolubov transformations are only relevant   in the right moving sector, and  the  corresponding  transformation from {\it in} to {\it exterior} modes is  what  accounts  for the Hawking radiation. As  is  well  known, the fact that  the initial   state,   corresponding  to  the   vacuum   for the right moving modes, and the    left moving pulse    forming  the black hole    $\kets{\Psi_{in}} = \kets{0_{in}}_{R} \otimes \kets{Pulse}_{L}$   can be  written as:
\begin{equation}
N \displaystyle\sum_{F_{nj}} C_{F_{nj}} \kets{F_{nj}}^{ext} \otimes \kets{F_{nj}}^{int}\otimes \kets{Pulse}_{L}, \label{inst}
\end{equation}
where the states { $\kets{F_{nj}}$}  are characterized by the   finite occupation numbers  {$\lbrace F_{nj} \rbrace $} for each  corresponding mode $n, j$, {$N$} is a normalization constant, and the coefficients {$C_{F_{nj}}$}'s are determined by  the Bogolubov transformations.
If one  now  decides  to  ignore the  degrees of  freedom (DOF) of the  quantum field lying in the   black hole  interior, and  to  describe   just  the  exterior  DOF,  one   passes, as usual, to a density matrix  description. Thus,    we   would obtain the reduced  density matrix   by    tracing   over the interior   DOF,   and we  end  up,   as   is   well known,  with  a  density  matrix  corresponding to  a thermal  state.   Note that  this density matrix   represents   in the language of \cite{dEspagnat, timpson}, an {\it improper} mixture  as it arises from  ignoring part of the system,  which as  a whole  is in a  pure state.  We  will thus  say that  what we have   at this point is an {\it improper thermal state}. As we said before, in this paper we will obtain a {\it proper} thermal state.

\subsection{CSL evolution of quantum fields in CGHS background}\label{subsec:op}

We are  now in a position  to show  how  a  thermal state,   corresponding to a \emph{proper}  mixture,  is obtained using our  adapted   version of CSL, and  some  reasonable  assumptions  about  QG.
  As the CSL  theory  involves  a  modification of the time evolution  of the quantum states,  we  need   a foliation of our space-time   associated  with a  ``global time parameter". As  is  customary in QFT,  we  will be  using an {\it interaction-type picture}, where  the  free  part   of  the evolution  is  encoded  in the  field operators,   and the interaction, which in our case is  just the  new  CSL part,   drives the evolution  of the states.    In  a   relativistic  context,    {based  in a  truly covariant version of CSL}, one would be  using a Tomonaga-Schwinger type of  evolution equation.

\begin{figure}
\centering
\includegraphics[scale=0.30]{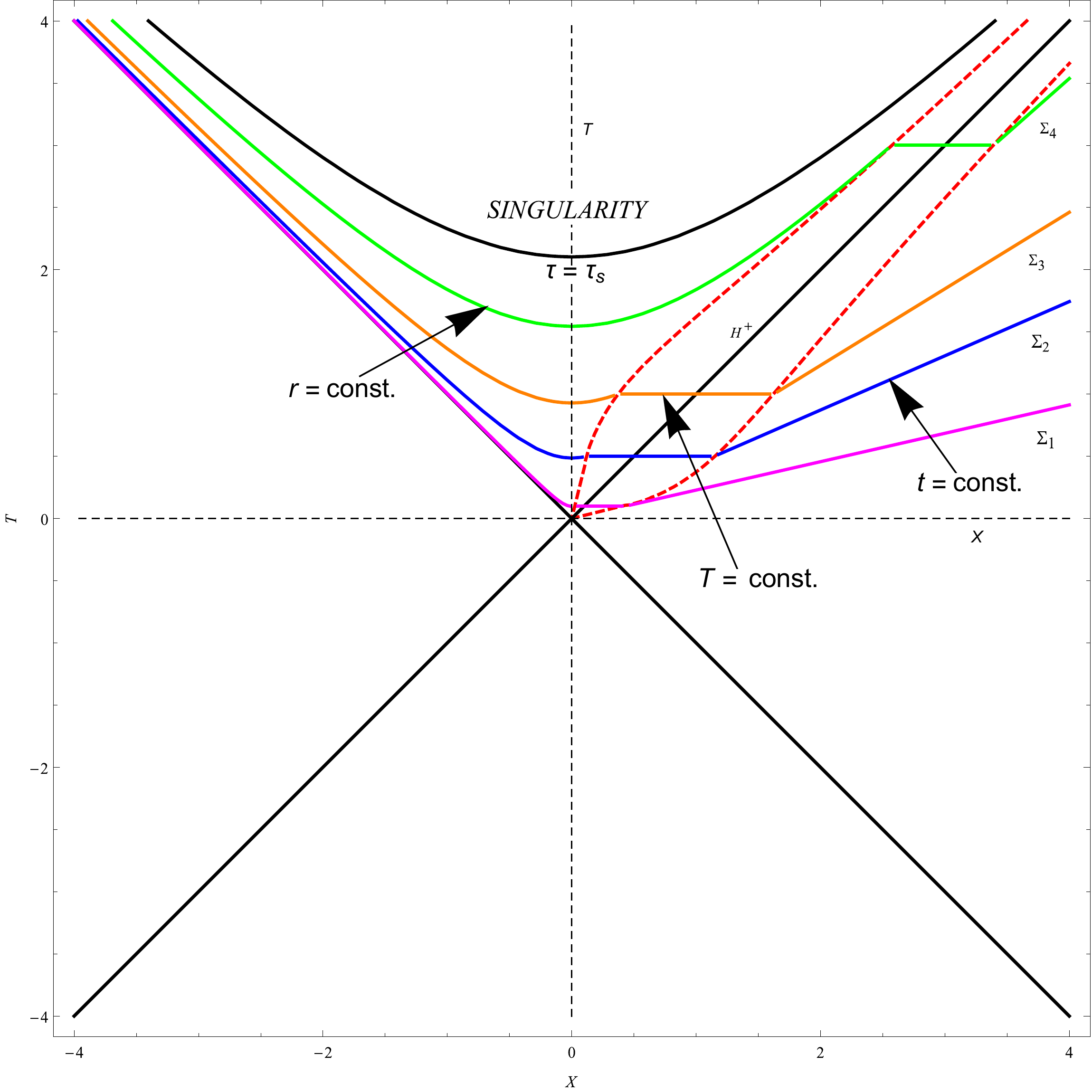}
\caption{Spacetime foliation plots for the CGHS spacetime.}
\label{fol}
\end{figure}

Let us now briefly comment on the  application of  CSL  to  the  CGHS  model.  In order to describe  the CSL evolution we   first need to foliate the spacetime with Cauchy slices\footnote{ We  must  be  careful here  as the  space-time  that includes  the QG region (i.e. the would be singularity)  can  not   be said to be  globally hyperbolic.   However, the space- time  to the past of  that region is, in fact,   globally  hyperbolic,  and there the notion of Cauchy  slice  is  appropriate.} which are defined in the following manner. We choose a {$r=const.$} and  a {$t=const.$} surfaces in the inside and outside of the horizon, respectively, and   then join them  using  a  surface { $T =const$} (see Figure \eqref{fol}). We specify the juncture  points (i.e., the points of intersection between $r=const.$ with $T=const.$ inside the horizon as well as $T=const.$ and $t=const$ outside the horizon) by  two curves,   $T_{1,2}(X)$. We take $T_1(X) = \left(X^2 + \frac{M}{\Lambda^3}e^{-2\Lambda/\sqrt{X}}\right)^{1/2}$ and $T_2(X) $  is found by reflection of  $T_1(X) $ with respect to the horizon $T=X$. Note that this choice for the intersection curves is  not unique and many other choices are  acceptable. However it is  essential  that the slices be  Cauchy  hypersurfaces.
With this construction,  we  can now introduce a ``global time parameter'' {$\tau$}  specifying   the   hyper surfaces of the foliation  by the value of   its  intersect with the {$T$} axis. This  foliation  covers the  complete  ``BH  region" (exterior and  interior) and one can extend it to  past vacuum region arbitrarily with no effect in our study, simply  because, the CSL modification in the dynamics  is significant only inside the black hole (more precisely, close to the singularity).

Now, we  must select the   operators  driving  the  collapse  in our version of CSL evolution. We  note, as   was already  mentioned,   that the CSL equations can be generalized to drive  the collapse process leading to one  element  of the  joint  eigen- basis of the set of  commuting operators {$\lbrace A^\alpha \rbrace$ } which we call the  \emph{collapse operators}.
 For each {$A^\alpha$} there is   one stochastic  function {$w^\alpha (t)$},
and  the  evolution equation 
takes the form: 
 \begin{equation}
 { |\psi,t\rangle_w=\hat {\cal T}e^{-\int_{0}^{t}dt'\big[i\hat H+\frac{1}{4\lambda}\sum_\alpha[w^\alpha(t')-2\lambda\hat A^\alpha]^{2}\big]}|\psi,0\rangle.} 
 \label{csl-ev}
\end{equation}
In standard non-relativistic CSL on flat spacetime, and as  previously noted,  one takes these  to be  the  (smeared) position operators so that the  collapse takes place in "position space"  so as to literally localize the wave function. In the situation at hand  we  will use, for simplicity,  the particle  number operators   (see footnote 3), so that, given that  $ \lambda$  becomes large  only in  regions of   high   space-time  curvature
the  states will collapse  to a state with  definite number of particles  in the  \emph{inside}  region.
 Far from the singularity the rate of collapse will be much smaller, and the  direct  effects of CSL evolution will be almost  negligible (i.e. just  as the ones  of  the original version of CSL).

We  should note  here that, we have  selected these  collapse   driving operators  for  simplicity of the  calculations but  other   choices   can be expected to
 yield  essentially  the same   end result. 
 The  basic  reason  for that is that  the relevant collapse operators will be associated  with the regions near the curvature  singularity  which  lies  inside the horizon,  and   the fact  that  CSL theory is known  not to  lead to   faster than light communication in  EPR-B type situations \cite{NLcollpase}.
 We  can understand  this  by  recalling that even  in  simplistic  versions of theories involving  measurement-related collapse, such as  the Copenhagen interpretation of quantum  mechanics,  for an EPR-B  situation,  the   choices  made  by Bob on  what  observable to measure in  his part of the entangled  pair,  can not be  used to  send  a signal to Alice,  who  will, for all the  possible  choices    made  by  Bob, observe the  same   statistical results in all  her possible observations. As we know, it is   only by measuring correlations between the two  sides,  that     something nontrivial   can be said in such cases.   In other words,   the   observers  outside  will  find   the same statistical  distribution of results for all  possible observations  regardless of  what  ``nature  measures'' (by this  we mean  ``in   what   basis,  does  the fundamental CSL  type theory  actually collapse our   quantum  field states'')  in the  black hole interior.

As we have mentioned before, our version of CSL in curved space-time is taken to have a \emph{curvature   dependent}   $\lambda$. Concretely,    we  will assume   here  that {$\lambda$}  depends, in this  2-dimensional  situation, on the Ricci scalar  $R$ according to:
\begin{equation}
{\lambda (R) = \lambda_0 \left[1+\left(\frac{R}{\mu}\right)^\gamma\right]} \label{lamr}
\end{equation}
where 
{$\gamma \geq 1$} is  a  constant, {$\mu$}  provides  an appropriate scale and $\lambda_{0}$ is the standard CSL rate of collapse \cite{comment}. It should be noted that, in a more realistic 4-dimensional model one should have a $\lambda$ depending on Weyl scalar ($W_{abcd}W^{abcd}$) as mentioned in \cite{Okon:2013lsa}. 

Note that the hypersurfaces given by our chosen  foliation in Figure \eqref{fol}  have constant {$R$} inside the black hole (in almost all the part of {$\Sigma_\tau$} that lies inside) and the value of $R$ is increasing towards the singularity. As a result, Cauchy slices with  larger $\tau$  have, inside the horizon, a larger  value of   $\lambda$ (inside the horizon   we have $R = \frac{4M\Lambda}{M/\Lambda - (T^2-X^2)} = \frac{4M\Lambda}{M/\Lambda -\Lambda^2 \tau^2}$). Therefore, although the CSL evolution  allows  us to evolve  the state of the quantum field from one Cauchy slice to another, the effective collapse will take place only in the interior part of the slicing, while its effect on the exterior portion will not be significant. Of course, there will be an indirect effect to the quantum field states defined with respect to the exterior Fock space, since these states are entangled with the internal field states. When the collapse  takes place for  the  inside  DOF, to the eigenstate of number operator defined inside, it will automatically collapse  also the exterior quantum states to a definite particle number, but that, of course, does not affect the statistical results obtained by the outside observers as explained before.

Thus, for the region of interest we have {$\lambda=\lambda(\tau)$} and  the resulting evolution will  achieve in   the finite  ``time''  to the  singularity (i.e., $\tau=\tau_s = \frac{M^{1/2}}{\Lambda^{3/2}}$),   what   ordinary CSL  achieves in an infinite  amount  of time.
  In this time the CSL  evolution   will drive the state  to one of the eigenstates of  the collapse  operators\footnote{We  noted that   as  we will be  working in the interaction picture,  we  must    make  the replacement   { $\hat H \to 0 $}  in \eqref{csl-ev}.}.
 Recall that the particle content of state { $\kets{F}$} is given by the particle distribution {$F=\{\dots, F_{nj}, \dots\}$} where { $F_{nj}$} is the number of particles in mode {$v_{nj}$} (both  for the inside of the black hole and the outside).  It should  be  noted that the collapse process  avoids   generating  singularities in the energy-momentum tensor when  we take  smeared   versions of these  number operators. That  ensures that  no ``firewall" type  situations  would arise far from the singularity\footnote{The issue  here is related to the fact that  measuring   the exact  number of  Unruh  particles present  in the  Minkowski  vacuum   would lead to the kind of singular   quantum states    that contain  firewalls. Such  eventuality  is   prevented   by   assuming that  the effects  of  all physical  devices  associated with  the measurement  can be represented by   smooth   local operators \cite{wald-pc}.}. The action of the number operator { $N^{int}_{nj}$} acting on { $\mathscr{F}^{int}$} is
$
{\hat N^{int}_{nj} \kets{F}^{int}= F_{nj} \kets{F}^{int}}.
$
 The set of collapse operators we are contemplating are:
\begin{equation}
{\hat {A^\alpha} =\hat{ N^{int}_{nj}} \otimes \mathbb{I}^{ext}} \label{csl-op}
\end{equation}
for all { $n, j$}, where {$\mathbb{I}^{ext}$} is the identity operator in  {$\mathscr{F}^{ext}$}. Thus the quantum states corresponding to interior Fock space will tend to collapse to the joint  eigenstates of $\hat{ N^{int}_{nj}}$ with definite  particle content, whereas states in the exterior Fock space will remain   essentially unaffected by  the collapse dynamics. As we said, the only  significant  effect on external states will be that due to the entanglement with internal quantum states of the black hole.

With the above construction we are now in  a position to show how quantum
evolution takes place in our model. We will do that, first, for a single system  (which is given in
\eqref{inst}), and  later,   for  an ensemble  of systems prepared in the same initial state.

\begin{figure}
\centering
\includegraphics[scale=0.35]{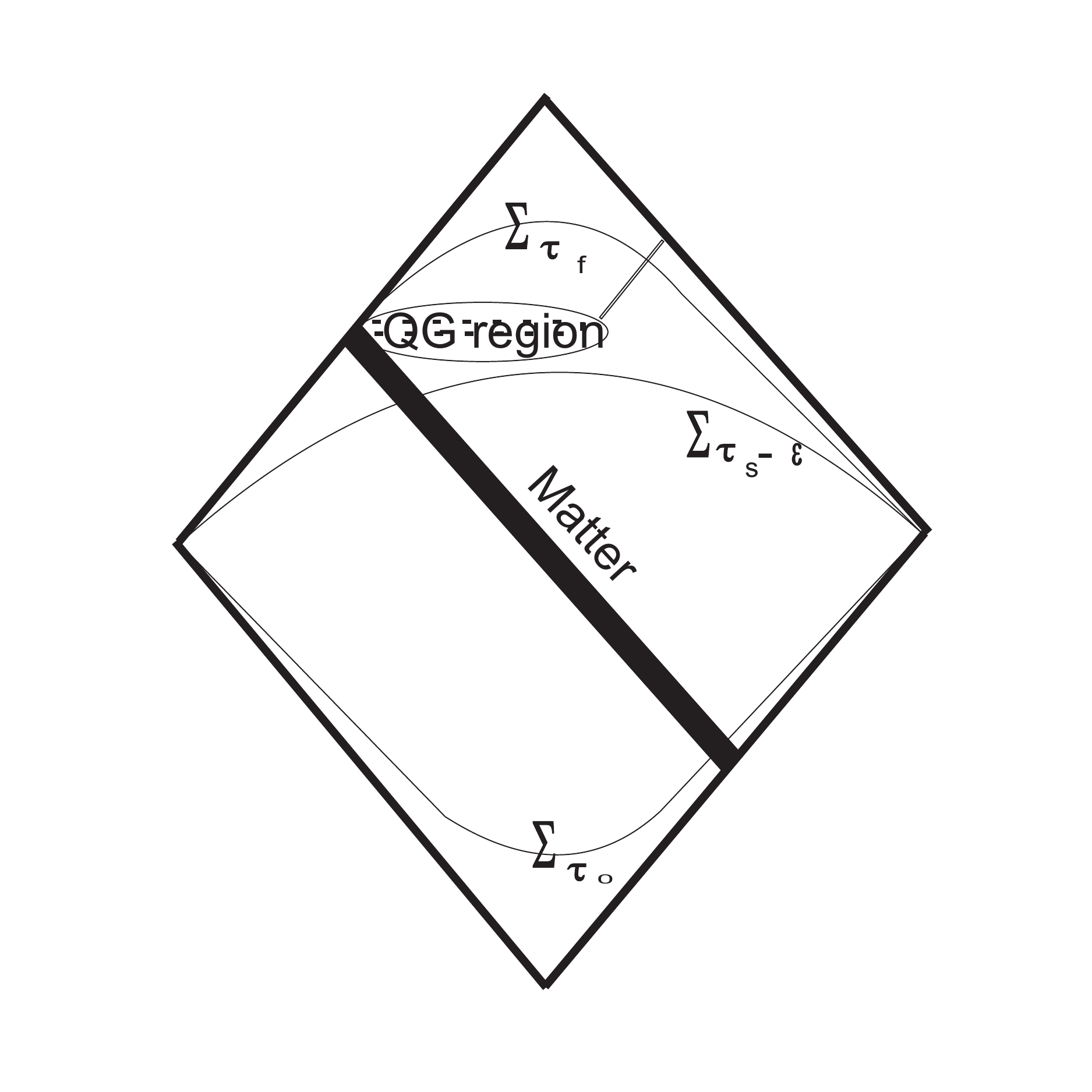}
\caption{Plausible space-time structure  including  the post-singularity  region.}
\label{finalst}
\end{figure}

Let us consider the first case. The fact that  CSL evolves   states towards one of the  eigenstates
of the collapse operators  \eqref{csl-op}  ensures that,  as  the result of  the  evolution \eqref{csl-ev},  the   state  at  a hypersurfaces {$\tau = const.$}, that comes very close to the  singularity (on $\Sigma_{\tau_s - \epsilon}$ in Fig. \ref{finalst}), would   be  of the form:
  \begin{equation} \label{hadamard}
{\kets{\Psi_{in, \tau}} =  N \displaystyle C_{F_{nj}} \kets{F_{nj}}^{ext} \otimes \kets{F_{nj}}^{int}\otimes \kets{Pulse}_{L}}.
\end{equation}
Note that  there  is no  summation,  so the  state is pure, even though it is  undetermined,  simply because  we  don't know  the actual realization of the stochastic functions $w_{nj}$.

Note that,   although  individual states with  definite occupation  number  in the  {\it ext} and  {\it int}  modes such as  the one above   lead to  singular
$\langle T_{ab}  \rangle$,   those  states  are  only  approached   asymptotically as  $\tau \to \tau_s$. The dynamics of  CSL
 generates  only  smooth  states  prior to the  singularity \cite{ARXIVE}. The situation is  analogous to the measurement  of a precise  number of Rindler particles in the  Minkowski vacuum in a finite time,  which is   impossible unless   the  field-detector interaction is singular  \cite{wald-pc}.

\section{Role of quantum gravity and the final result}
We have characterized the state after the CSL evolution \eqref{hadamard} defined on a Cauchy
slice  that comes   very  close to  the singularity,  and  thus we must  now consider  the  role of quantum gravity,  to pass to the final hypersurface $\Sigma_{\tau_{f}}$ in Fig. \ref{finalst}. For that, we shall make some  seemingly  natural
assumptions about QG. We will  assume that  a  reasonable  theory  of QG     will resolve the singularity  and   lead,  on  the other  side, to   a reasonable space-time. Moreover, we  will assume that such a theory  will not lead to  large violations  of the   basic  space-time conservation laws.
  With these  assumptions,  we  now  look  at the situation  on the region   just   before the  singularity:  There, from the energetic  point of view,  we   have  the  following:
  {i)}  The   incoming   positive energy flux corresponding to the   left moving pulse  that  formed the  BH.
   {ii)} The incoming flux (from region II to region III in Fig. \ref{cghs}) of   the  left  moving   vacuum  state for the rest of the modes  which is known to  be  negative  and  essentially   equal to the total Hawking Radiation flux.
  { iii) }  The flux associated  with the  right moving modes that crossed  the   collapsing matter (from region I' to III) but  fell  directly into the  singularity.

The only energy  missing in the  budget  above is that    associated  with the   Hawking radiation flux.  If energy  is to be  essentially  conserved  by  QG,  it seems that the only  possibility  for the  state    in the  ``post singularity region"    is  one corresponding to  a very small value of the energy. Such state
of  negligible energy  might be  associated with  some   remnant  radiation, or  perhaps a compact and stable remnant  with   something close to   Planck's
mass.  Those  possibilities  would  represent  a case  were a  small  amount  of  information  survives  the  whole
process, and  will  be ignored   hereafter  for  simplicity.  Instead,  we   will  consider  the simplest  alternative:
A zero  energy momentum  state  corresponding to a ``trivial"  region of space-time.    We denote it by {$\kets{0^{p.s.}}$} (post-singularity).  In other words,  we complete  our
characterization of the  evolution,  by   assuming that  the effects of   QG can be represented  by the curing of the singularity, and by the  transformation:
        \begin{equation}\nonumber
{\kets{\Psi_{in, \tau}} =  N \displaystyle C_{F_{nj}} \kets{F_{nj}}^{ext} \otimes \kets{F_{nj}}^{int}\otimes \kets{Pulse}_{L}}
\end{equation}
   \begin{equation*}
{ \to  N \displaystyle C_{F_{nj}} \kets{F_{nj}}^{ext} \otimes \kets{0^{p.s.}}}.
\end{equation*}
The above result  might  seem  a  bit unsettling, because  we  end up  with a  pure quantum state,  while,  if  information is  lost  in the full  process, we  should  end  with a  mixed (thermal) density matrix.
The point, of course,  is  that   the theory does not   predict   which one of the possible pure states  we will end up with.  That  depends on the particular realization of the random functions { $w^\alpha$}    that   appears in the   CSL   evolution  equation. 

 Thus, as  a consequence of the inherent  stochasticity of  the  theory, all we can do  is to  make statistical  predictions.  In fact,   we can easily do  that by
 considering  the case where an ensemble of  systems are identically prepared   in the same  initial state:
\begin{equation*}
{\kets{\Psi_{in}} = \kets{0_{in}}_{R}\otimes \kets {Pulse}_{L} }.
\end{equation*}
We  then describe this   ensemble by the pure density matrix
${  \rho(\tau_0)  = \kets{\Psi_{in}}  \bras{\Psi_{in}}}$.
In  each case the pulse  will  lead to the formation of a black hole.
As before, CSL   evolution  is the dominant   where the  curvature becomes large,  i.e., before reaching the singularity, or  more precisely  the quantum gravity regime (i.e. the  hypersurface $\Sigma_{\tau_s-\epsilon}$).
The evolution   from  initial hypersurface {$\Sigma_{\tau_0}$} to an intermediate  hypersurface {$\Sigma_{\tau}$}  is  then given by
\begin{equation}
{\rho(\tau) = {\cal T}e^{-\int_{\tau_0}^{\tau}d\tau'\frac{\lambda(\tau')}{2}\sum_{nj} [\tilde N_{nj}^{L}-\tilde N_{nj}^{R}]^{2}}\rho(\tau_0)}.
\end{equation}
We express {$\rho(\tau_0)=\kets{0}^{in}\bras{0}^{in}$} in terms of the  {\it out} quantization (ignoring  left moving modes as they are not relevant for  the  external observers):
\begin{equation*}
{\rho(\tau_0) = N^2 \sum_{F,G} e^{-\frac{\pi}{\Lambda} (E_F +E_G )}  \kets{F}^{int}\otimes\kets{F}^{ext} \bras{G}^{int}\otimes\bras{G}^{ext}},
\end{equation*}
 where {$\Lambda$} is the parameter of the CGHS model and {$E_F \equiv \sum_{nj} \omega_{nj} F_{nj}$} is the energy of the  
  state  $\kets{F}^{ext}$  with respect to late-time observers near {$\mathscr{I_R}^+$}.

The operators {$\tilde N_{nj}$} and their eigenvalues are independent of {$\tau$}. Thus we have, 
\begin{equation}\nonumber
\rho(\tau) = N^2 \sum_{F,G} e^{-\frac{\pi}{\Lambda} (E_F +E_G )} {\cal B}
{\kets{F}^{int}\otimes\kets{F}^{ext} \bras{G}^{int}\otimes\bras{G}^{ext}},
\end{equation}
 where  ${\cal B} \equiv e^{- \sum_{nj} (F_{nj}-G_{nj})^2\int_{\tau_0}^{\tau}d\tau'\frac{\lambda(\tau')}{2}}$. In general, this equation does not represent a thermal state. Nevertheless, as
 {$\tau$} approaches the singularity, say at
 {$\tau=\tau_s$}, the integral
 {$\int_{\tau_0}^{\tau}d\tau'\lambda(\tau')/2$ } diverges since  {$\lambda(\tau)$}  becomes large on  hypersurfaces of high curvature (as we have assumed $\gamma \geq 1$ in \eqref{lamr}).
 Then, as
 {$\tau\to\tau_s$} the non diagonal elements of
 {$\rho(\tau)$} cancel out and we have:
\begin{equation*}
 {\lim_{\tau\to\tau_s} \rho(\tau) = N^2 \sum_{F} e^{-\frac{2\pi}{\Lambda} E_F}  \kets{F}^{int}\otimes\kets{F}^{ext} \bras{F}^{int}\otimes\bras{F}^{ext}}.
\end{equation*}
If we now  include  the  left  moving pulse,  and  take into account  what we  have  assumed  about    the role of QG,   we  obtain  a   density matrix  characterizing the    ensemble { \it after} the singularity (on $\Sigma_{\tau_{f}}$ in Fig. \ref{finalst}),  which is   given by
 \begin{eqnarray}
\rho^{Final } &=& N^2 \sum_{F} e^{-\frac{2\pi}{\Lambda} E_F} \kets{F}^{ext} \otimes \kets{0^{p.s.}} \bras{F}^{ext} \otimes \bras{0^{p.s.}}.\nonumber\\
  && \equiv  \kets{0^{p.s.}} \bras{0^{p.s.}} \otimes \rho^{ext}_{Thermal}
\end{eqnarray}
That  is,  we  started  with an ensemble  described by a  pure state  of the quantum field  corresponding to  the vacuum  plus  an initially  collapsing pulse,  and   the corresponding space-time  initial data on   past  null infinity,  and  ended  up, with a  \emph{proper}  thermal state. In other words,    we ended  up with   a proper mixture  (recall the terminology from \cite{dEspagnat,timpson}), characterizing an ensemble of systems  with a thermal  distribution on  future null infinity  followed  by an empty region.

\section{Summary and discussions}
We  have  studied,   at the quantitative level,  a  concrete  realization of  the  idea  that  a resolution of the   black hole information conundrum might   be related to  the solution of the  measurement problem in quantum mechanics.  We have seen how  a   proposal   developed  for addressing the  latter can be  adapted  to  the black hole  setting, and  that by  assuming that the   parameter
controlling the   departure from   the   usual quantum mechanical unitary  evolution  increases  with   space-time   curvature, we are  led to a picture  whereby the   information  loss,  one  can  naively  infer
as occurring  in the latter  case,  becomes  a direct and actual result of the modified  dynamics. This leads to a situation  where  the  information loss    associated  with the full evaporation of   a black  hole
via  Hawking   radiation   can be  seen,   not as a  paradoxical   result, but as  the natural  outcome  from the
fundamental  quantum   evolution that  accounts for the physics of both micro and  macro objects.

 Of  course,  we  assumed that a  QG theory would   resolve the singularity, and otherwise  be reasonable  so that   it  does not lead  to gross   violations  of  conservation laws,  with  potentially observable   implications  in   the    regions where   something close to a  classical space-time description is expected.

 We  should also acknowledge that in the course of this  work  we made    several  simplifying assumptions,  and ignored  some issues  that are not  easy to deal with, but we   think that  there are good  reasons to  expect   that the general picture we obtained  is  rather robust.   We next   deal  briefly  with the most important  of these approximations and issues:



1. {\it Choice of  collapse  operators}:  The  general   choice of   the  operators $A_{\alpha}$  controlling the CSL  dynamics  is  clearly an  open issue. As  we  already mentioned,  a complete theory  should specify,  for all possible circumstances,  and in a manner that depends only on  the  dynamics of the system  in question,   what determines  such operators.  It seems   that   they must correspond  to  a kind of smeared  position operators,  in the case  where  the  system can be described in terms of the non-relativistic  quantum mechanics of many particles.   The choice of the   particle number operator  in the interior  of black  hole  region   that    we made in  section \ref{subsec:op}, was clearly done for convenience. In a fully  covariant   theory  we  expect these operators to be locally constructed  from   quantum fields,  and the number operators  are clearly non-local.  It  should be  pointed  out, however, that  in  these  theories,  one
can rewrite  the same    CSL evolution   in terms of  various    and   very  different  choices of operators,   as is  in  fact  shown  explicitly  in    the work \cite{pedro}.
 On the other hand, as we already  noted in connection with   EPR-B situations,  the   collapse  of  the state into  eigenvalues of  an operator  associated  with a  certain  space-time region  has  no  influence  whatsoever, in the    effective description, in terms of a density matrix,   for the state  restricted to space-time regions that are causally disconnected.
   Of course,  in the case of an EPR experiment,  one could still consider the correlations  between the outcomes of Alice vs. Bob measurements, and those will clearly depend on what quantity  did each one measure.
    However,  in our case, the state of the subsystem corresponding to Bob (i.e. the matter field in the inside BH Region) will simply disappear. That is, in the  post  Quantum Gravity  region the state will correspond to a sort of vacuum state containing   almost  no energy or information,  so  that there would be (almost)  no other correlations to look for.
 We  fully  expect  such   robustness (\emph{i.e.},  independence  of the precise   choice of   collapse operators in the  region  III) to   apply to the  characterization of the state of the field  in  $\mathscr{I}^{+}_{R}$ in terms of a density matrix.

2. {\it Relativistic Covariance}:  The model we  have  employed  is  based on a  non-relativistic  spontaneous   collapse theory, and  it  is  clear that a  satisfactory proposal to deal with the issue  at hand  should  be  based on  a  fully covariant  theory.   We should note, in this  regard,  the
 early studies \cite{advances}, and the recent   specific proposals  for      special relativistic  versions of  these  type  of theories\cite{Tumulka-Rel,
Bedingam-Rel, RelColPearle}.
 In those  theories  the   evolution, in    the   interaction picture  approach we have  been  using,  should   be  describable in  terms of  a  Tomonaga-Schwinger evolution  equation\footnote{Recall that  the  Tomonaga -Schwinger equation  $ i \delta \kets{\Psi (\Sigma)}  = {\cal {H}}_I(x)\kets{\Psi (\Sigma)}  \delta \Sigma(x)$    gives the   change in the interaction picture for the state   associated  with the  corresponding    hyper surfaces   $\Sigma'$ and  $\Sigma$,  when the   former is  obtained  from the  latter  by    an infinitesimal   deformation   with  four volume
  $\delta \Sigma(x)$  around the  point $x$ in $\Sigma$. We are  also ignoring  here   the   formal aspects  that indicate that strictly speaking the interaction picture does not exist.}  where, instead  of the   interaction    hamiltonian,  we  would have the corresponding  collapse  theory  density operator.
 We  hope  eventually to  adapt the present scenario to those proposals,  a task we  expect to  be rather nontrivial.

3. {\it Choice of  foliation}: We  have  presented  the  analysis  using a very particular  foliation.
  Within  a  covariant  setting, we  can  expect  that any  specific   physical prediction   should be  independent of the foliation.
   One important consequence  would be  that,   whenever we  consider a  foliation  passing  through  a  region of space-time  which is  far  from the singularity, the  changes in the  state around that point,  associated  with the   CSL  type modifications,  should    have  effects in  local  operators   that   will  be  very  small,   simply because the   CSL-like  parameter  $ \lambda$  will be small there. This  indicates  that  we  should not encounter anything like ``firewalls" in the region of the horizon\footnote{Our   references  to ``the  horizon"  within the setting where the singularity has  been replaced  by the ``quantum gravity region",  should be  taken to indicate the   boundary of the past  domain of  dependence  of  the said  region.} which  is  far  from the singularity.

4. {\it Energy  conservation}:  One  might be concerned  that,  when considering an individual  situation,   the  energy of the initial pulse of matter  might not be  exactly equal to that corresponding to the  state   with definite  number of particles $\kets{F}^{ext}_R $ that characterizes the  modified matter content  in the asymptotic  region,  once the    black hole    has  evaporated  completely.  The first thing to note is that CSL,  in general,  leads
to small violations of energy conservation   and that issue,  in fact,  has led
to  the establishment the  most  stringent  bounds  on the parameter  $\lambda$ (although modified  covariant theories might   evade  this problems
altogether.  See  for instance  the detailed  discussions in  \cite{Bedingam-Rel,  Unruh-Wald}). The  second thing  to note is that if there is  small
amount  of energy  remaining inside the  black hole region,   and very close
to the singularity,   simply because the positive   and negative energy fluxes
do not cancel  each other exactly,  there would seem to  be  no problem, at
least  in principle, if  such  energy is radiated   after  the singularity.  In that
way, most of the  initial  energy  will be radiated in the standard  Hawking
radiation,  and  a   very small amount of  energy,  carrying  a minuscule
amount of information,   would remain to  be radiated  towards infinity  from
the ``quantum gravity  region".


5. {\it Back reaction}:
 We  must  note that,  in  all the   discussion  so  far, we  have omitted  the   very important  issue of  back reaction.  The  changes  in the  space-time  metric as  a response to those  in state  of
 the matter  fields,  are essential, if we want to  account   for the decrease of the  mass of the  black hole  as a  result of the  Hawking radiation   taking  energy to infinity.
 That,  in turn, is  an  essential aspect of the   arguments involving  overall  energy  conservation.
 Going further,   the change in the  space-time metric,  and,  in particular, in  the  black hole's mass, and  its ``instantaneous  Hawking temperature" in a   more realistic  model, are expected to  modify the  nature of the radiation,  so  that, the ``late  time"  radiation would be, in a  sense,   emitted  at a higher temperature  than
 the ``early times" one,  leading to the runaway effects  that are  associated  with  the expectation of an
 explosive  disappearance of the  black hole  itself.  It is  clear that all  such  effects  are  extremely important  in
  obtaining a  realistic  picture of the entire   history of formation   and  complete  evaporation of a black hole.
 However,  there seems to be no reason to  expect that  those important  changes  will modify, in  an  {\it essential
 way}, the workings of the proposal  we have been considering  here.
In fact, the   back reaction   is  essential,    in accounting   for the decrease of the Bondi mass
to  a  very small value, as  one  considers the  very late  parts  of $\mathscr{I}^{+}_{R}$.
This,  in turn,   is  what  allows  us to consider,   matching  the asymptotic  region   with  the space-time that  is
expected to  emerge  on the other side of the  quantum gravity region that replaces the ``would be  singularity",  and which, as  we have  indicated,  should be  thought as essentially  empty   and flat, (with the caveats  discussed  in the  item  above).

On  the other hand,  it  seems that   theories  involving   spontaneous  reduction of the   quantum  state  of
matter fields should   facilitate, at least  at the conceptual  level,  the  treatment of  back reaction\footnote{ In
the standard  treatments  there seems to  be an ambiguity  regarding  the use of  ``in-in"   or ``in-out"
expectation values \cite{in-out}.},  simply  because  there is a clear state  that should  be  used  in  evaluating
the expectation value of  the energy momentum  tensor  at  each hypersurface:  in our case the state
associated  to  that  hyperurface by the CSL dynamics.    Of course, as  indicated before,    a fully satisfactory   account
would have to  rely on a  fully covariant theory   corresponding  to   general relativistic  generalizations  of  those recently proposed  in\cite{Tumulka-Rel,
Bedingam-Rel, RelColPearle},  and,  as  discussed in  those works,   the appropriate  state  to  be used in computing the
expectation  value  of  an operator associated with a given  space-time  ``event" $p$,   would  be  the state  associated  with the past
null cone of $p$ (more precisely one  would need to consider open regions  around  $p$ and   the  state  associated  with  the boundary  of the causal past  of those). Nevertheless,  and  despite the  discussion  below, it is clear that the  technical  aspects of the treatment of back reaction  need  to  be  further  analyzed  and developed  (particularly in view of  the  footnote  10 below,  and  related  comments) if one  wants to have a viable,  even if  only approximate,  semi-classical  account of the  black hole evaporation process.

6. {\it Reliance  on semi-classical  gravity}:  The previous   item  forces  us   to consider  the  basic  viability of   semi-classical   gravity, the scheme where one treats the  space-time metric  at the classical level,  but uses as  a  source in  Einstein's  equation  the quantum   expectation value of the  energy momentum tensor. This  question   has  been  considered   in  a well known  paper by Page and  Geilker \cite{Page}.  That  work  is often referred  to, as  indicating that  semi-classical gravity   is  simply at odds  with experimental results.  However,  what is not often  noted is that  such conclusion is only valid  in the contexts  where quantum mechanical evolution   does not  include  any sort of   measurement-related or spontaneous  collapse.  Thus it  certainly  would not be relevant  for  our proposal.  On the  other hand,   it is  clear that if  one   wants  to incorporate  the   reduction or  the collapse of the  quantum state  of  matter fields,    the   semi-classical equation, taken   to  be   fundamental  and  100\% accurate,  would   not   be viable,  as it is  simply inconsistent\footnote{The fact is that   $\nabla^a \langle T_{ab}  \rangle  \not= 0 $  during  the  collapse  of the  quantum state,  so $ \langle T_{ab}  \rangle  $  it  can not  be equated  with $G_{ab}$, which   identically  satisfies  $ \nabla^a G_{ab}=0$ . }. The point however is that if one  takes   the   metric  description  of  space-time  as  merely  an effective and approximate  characterization,  in analogy, say,  with Navier -Stokes  (NS) equation in hydrodynamics,  one  would not expect  the  equation to hold   exactly,  or to be  valid  universally.    That is,  just as there would  be situations  where   the NS  would not be  an appropriate  characterization of  the   fluid,  such as when  an ocean  wave  breaks, and   in which  one can expect  important  local departures from the equation,   one   can  expect   something analogous  to  occur  with the  semi-classical   Einstein  equation.  This,  however,  would not  invalidate  the  equation in its use  for   some  suitable  macroscopic  and  approximate  characterizations.  More precise  characterizations, both in the  treatment  of fluids  and  of   gravitation,  can be expected to involve higher  order and more   complicated terms,  and,  of course   eventually,   as the  natural  scale of the   more fundamental and   underlying  theory  is approached,   one  would expect the complete  breakdown of the  effective   description.    Some initial steps in the exploration of  the  formal  adaptation of this approach  to the use of   semi-classical  gravity  in a cosmological context  have  been considered in \cite{Alberto}.  We should also  note the  work  \cite{Carlip} in which the  general arguments against  semi-classical  gravity were critically considered,  concluding that  they  are not as  robust  as  it might have  seemed initially.

Finally,  it is worth  mentioning a  rather speculative, but very suggestive  point   made in \cite{Okon-New}. The
general  picture that  one obtains  in this  way of dealing with the BH information question has  the  kind  of  intrinsic  self consistency
of a boot-strap  model:  if  black hole  evaporation is  associated  with  loss of information and departure from quantum   unitarity,  then, it  is  natural, to  consider that virtual, microscopic  black holes, generated  "off-shell" in  radiative  quantum processes\cite{George},   should  themselves lead to loss of information and  departure of unitarity {\it in all  situations},  and that  could, perhaps,   be  the source of the effects  which one  parametrizes,  at the  effective level,  as   the   modifications of quantum mechanics  represented  in CSL  and related
theories.  The tail-baiting snake picture, would then,  be complete.

We reiterate that, at this point, this  work  represents  just  a toy model, as, in particular, all  items above   would  need to  be addressed  in any realistic version. However,   we  believe that  reasonable  models  with the  basic features we have discussed  here  do offer  a  rather  interesting path   to resolving the long  standing conundrum  known as  the ``Black Hole Information Loss Paradox" and,  to do  so in connection with   the  attempts to resolve,  what  we, and  various   other colleagues,  but certainly not the majority,  regard as  a very  unsettling  aspect of  our  current  understanding of quantum theory:   the  ``measurement  problem''.


{\it Acknowledgment:} We thank Robert Wald, Philip Pearle and Elias Okon for discussions. 
DS is supported by the CONACYT-M\'exico Grant No. 220738 and UNAM-PAPIIT Grant No. 
IN107412. SKM and LO are supported by DGAPA fellowships from UNAM.


\begin{thebibliography}{99}
\bibitem{comment} The original CSL theory was developed as a modification of non-relativistic many particle quantum mechanics and the version we are envisioning would be a formulation written in a field theoretical language.
\end{thebibliography}
\end{document}